\documentclass[preprint]{elsarticle}

\usepackage{lineno}
\usepackage{hyperref}
\usepackage{tabularx}
\usepackage{booktabs}
\usepackage{dblfloatfix}
\usepackage{multirow}
\usepackage{amssymb}
\usepackage{xcolor,color,soul}
\modulolinenumbers[5]
\journal{Journal of \LaTeX\ Templates}

\newcommand{\MP}{{\ensuremath{\textrm{C}^{}_{17}\textrm{H}^+_{11}}}}
\newcommand{\pyrch}{{\ensuremath{\textrm{PyrCH}^+_2}}}
\newcommand{\pyrc}{{\ensuremath{\textrm{PyrC}^+_7}}}
\newcommand{\bzium}{{\ensuremath{\textrm{C}^{}_{7}\textrm{H}^+_{7}}}}
\newcommand{\CCHH}{{\ensuremath{\textrm{C}_2\textrm{H}_2}}}
\newcommand{\CM}{{\ensuremath{\textrm{cm}^{-1}}}}

\begin{document}

\begin{frontmatter}

\title{Identification of the fragment of the 1-methylpyrene cation by mid-IR spectroscopy}
\author[airap,auzk]{Pavol Jusko}
\author[alcpq]{Aude Simon}
\author[airap]{Gabi Wenzel}
\author[anj,auzk]{Sandra Br\"unken}
\author[auzk]{Stephan Schlemmer}
\author[airap]{Christine Joblin\corref{cor1}}

\cortext[cor1]{christine.joblin@irap.omp.eu}

\address[airap]{Institut de Recherche en Astrophysique et Plan\'etologie (IRAP), Universit\'e
de Toulouse (UPS), CNRS, CNES,
9 Av. du Colonel Roche, 31028 Toulouse Cedex 4, France}
\address[auzk]{I. Physikalisches Institut, Universit\"at zu K\"oln,
Z\"ulpicher Str. 77, 50937 K\"oln, Germany}
\address[alcpq]{Laboratoire de Chimie et Physique Quantiques LCPQ/IRSAMC,
Universit\'e de Toulouse (UPS) and CNRS, 118 Route de Narbonne, 31062 Toulouse, France}
\address[anj]{Radboud University, Institute for Molecules and Materials, FELIX Laboratory,
Toernooiveld 7c, 6525 ED Nijmegen, The Netherlands}

\begin{abstract}
The fragment of the 1-methylpyrene cation, \MP, is expected to exist in two isomeric forms,
1-pyrenemethylium \pyrch\ and the tropylium containing species \pyrc. We measured 
the infrared (IR) action spectrum of cold \MP\ tagged with Ne using
a cryogenic ion trap instrument coupled to the FELIX laser. Comparison of the
experimental data with density functional theory calculations allows us to identify the
\pyrch\ isomer in our experiments. The IR Multi-Photon Dissociation spectrum was also recorded
following the \CCHH\ loss channel.
Its analysis suggests combined 
effects of anharmonicity and isomerisation while heating the trapped ions, as shown by 
molecular dynamics simulations.
\end{abstract}

% Molecular dynamics calculations were used to support an interpretation in which the 
%dissociation involves complex isomerization pathways with several intermediates.
%The IR Multi-Photon Dissociation 
%spectrum was also recorded by following the C$_2$H$_2$ loss channel but was found to 
%be difficult to analyse. Molecular dynamics calculations were used to support an 
%interpretation in which the dissociation involves complex isomerization pathways with 
%several intermediates.


\end{frontmatter}

%\linenumbers

%\section{Highlights}
%- $\MP\cdot \textrm{Ne}$ complexes are formed in-situ in a cold 22 pole ion trap via 
%ternary attachment.\\
%- The vibrational spectrum of the  $\MP\cdot \textrm{Ne}$  weakly-bound complexes is 
%recorded via infrared pre-dissociation action spectroscopy.\\
%- The structure of \MP is  identified as 1-pyrenemethylium \pyrch\ using DFT B3LYP 
%calculations.

\section{Introduction}
\begin{figure}[h]
\centering
   \includegraphics[width=.50\textwidth]{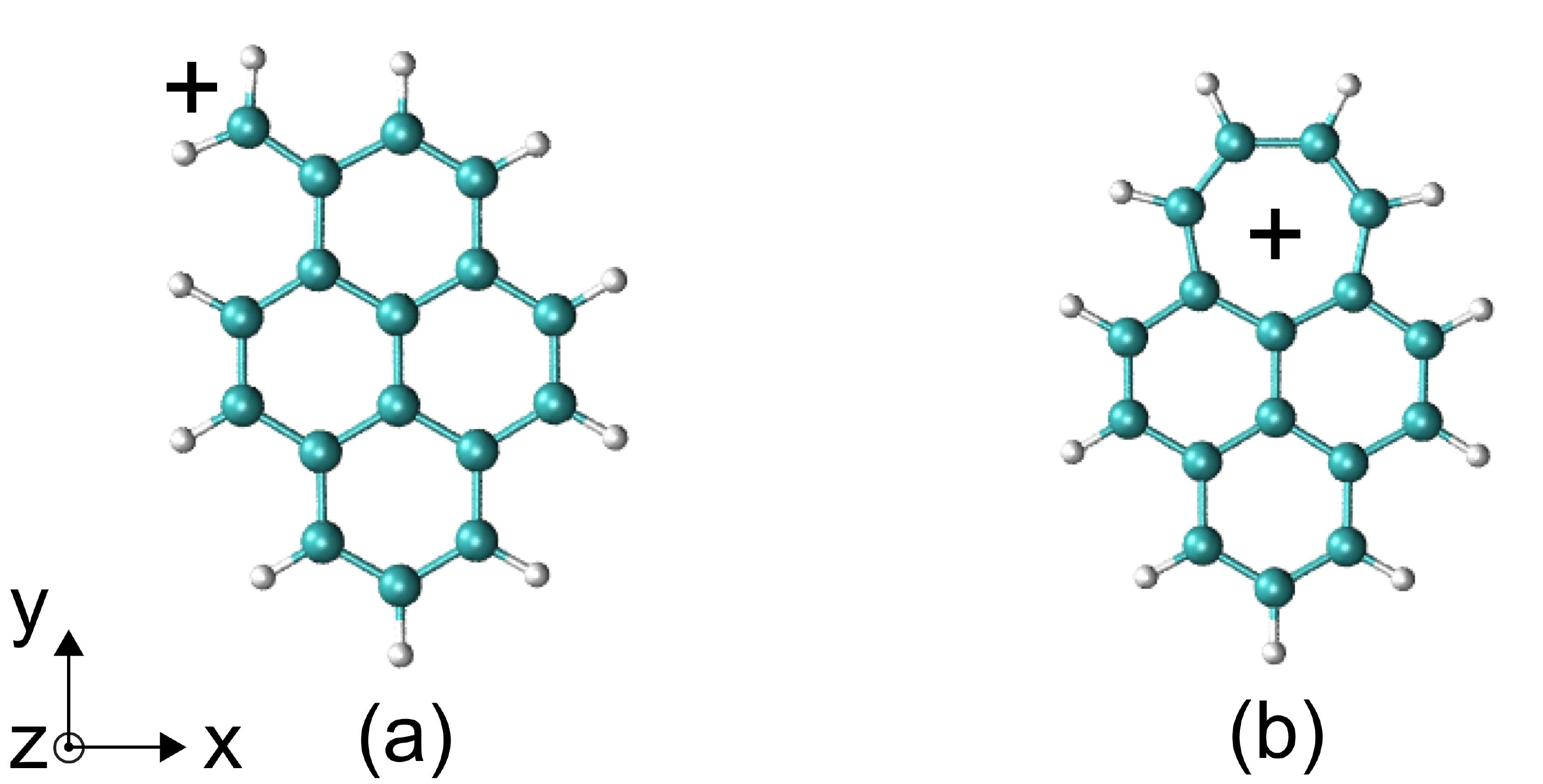}
  \caption{
  Topological formulas of the two isomers of \MP:
  \pyrch\ (a) and \pyrc\ (b).
  }\label{f:iso}
\end{figure}

The photoproduct of the 1-methylpyrene cation, \MP, has been pointed
out in matrix-isolation studies as a candidate of interest to account for some of the diffuse
interstellar bands including the strongest one at 4428\,\AA\ \citep{Leger1995}. This has
motivated experimental studies using the ion trap setup PIRENEA in which 
CH$_3$-C$_{16}$H$_9^+$ ions were trapped, irradiated by UV-VIS photons and its 
photofragment,  \MP,  could be isolated. First multi-photon dissociation (MPD) spectra 
were obtained in order to reveal the structure of these ions \citep{Kokkin2014}.
Rapacioli et al. performed extensive calculations on the possible isomers 
and isomerisation pathways \citep{Rapacioli2015}. These studies have shown that both the
1-pyrenemethylium isomer, \pyrch, and the isomer containing a tropylium cycle, \pyrc\ (see
Figure~\ref{f:iso}), could be present under the experimental conditions used in
\citep{Kokkin2014}. Both isomers were found to be quasi degenerate (cf. Table~3 in \citep{Rapacioli2015}).
In addition, theoretical calculations showed several isomerisation pathways 
with barrier heights in the range of $3.5-4.0$~eV \citep{Rapacioli2015}. This, together with 
the fact that both isomers coexist at formation, could explain why the recorded bands in
MPD measurements were so broad.  

Infrared (IR) spectroscopy is an obvious technique to assign
absorption bands to the \pyrch\ and \pyrc\ species and thus to disentangle the
nature of the 1-methylpyrene cation photoproduct.
Similarly to the case of the electronic transitions described
above, absorption of multiple photons at IR wavelengths can be used to probe IR transitions
of stored ions.
This technique is well-known as IRMPD spectroscopy and has been successfully applied to 
different types of molecules and molecular complexes including polycyclic aromatic 
hydrocarbons (PAHs), which are of interest for this study \citep{Oomens2006, Lorenz2007}. 
A drawback of the technique is that it implies the heating of the trapped ions by successive
IR photon absorption. As a result, anharmonic effects, which lead to modifications of the IR
spectrum with internal temperature, can therefore induce strong deviations of the recorded 
spectra compared to linear IR absorption spectra \citep{Parneix2013}.
In a case like the \pyrch\ and \pyrc\ species, an additional
complication can arise from isomerisation processes induced by the heating.

The recent development of a cryogenic trap instrument \cite{BEAMLINE} at the 
FELIX\footnote{Free Electron Lasers for Infrared eXperiments, \url{www.ru.nl/felix}} 
%(Free Electron Lasers for Infrared eXperiments) 
Laboratory \cite{Oepts1995} opens new perspectives for this study.  Indeed, it allows to use
a rare gas tagging often referred to as messenger technique which provides information on 
the IR spectrum of the cold ion--rare gas complex upon the absorption of a single photon.
This type of spectra was earlier reported in molecular beam experiments, in particular on
jet-cooled complexes of cationic naphthalene-Ar, phenanthrene-Ar and
phenanthrene-Ne \citep{Piest1999, Piest2000}. The possibility to build and store such
complexes in a cryogenic ion trap opens new perspectives \cite{Bruemmer2003,Jasik2013}, 
especially to isolate and study fragments of the parent species.

In this article, we present the first IR spectrum of the Ne tagged \MP\ species
together with the IRMPD spectrum of the bare cation. The analysis of the experimental 
spectra involves both static Density Functional Theory (DFT) calculations to interpret 
the cold IR spectrum and molecular dynamics simulations to gain insights into the hot 
IRMPD spectrum.

\section{Methods}
\subsection{Experimental}

The experimental spectrum has been recorded on the FELion beamline at the
FELIX \cite{Oepts1995} Laboratory at the Radboud University in The Netherlands.
The FELion instrument at the beamline \cite{BEAMLINE} consists of a cryogenic 22-pole 
rf ion trap equipped with two quadrupole mass filters allowing to mass select and analyse
the ions to be studied, as well as the products created in the trap. 
%An example of the recorded mass spectrum is given in Figure S1 of the Supporting Information (SI).
The ions of interest were created in a storage ion source \cite{Gerlich1992} via
electron bombardment of the 1-methylpyrene neutral gas evaporated directly
into the UHV chamber from a solid 1-methylpyrene sample (Sigma-Aldrich).
The sample was heated to a temperature of $40^{\circ}\,\textrm{C}$ in order to achieve
a sufficient vapour pressure. Several thousands ions of mass-selected \MP\ cations
($m/z=215$, cf. Figure~S1 in the Supplementary Information (SI) for the different experimental 
conditions used in this work) were stored in the trap for 
each experimental cycle, while
one of the following two spectroscopic techniques was applied:
\paragraph{IR Pre-Dissociation (IR-PD) of the weakly bound complex with Ne} The trap was
  	operated at a nominal temperature
    %\footnote{For the actual ion temperature determination  cf. \cite{BEAMLINE}
    of $8-9\;\textrm{K}$. Upon ion injection
    high quantities of  a Ne:He 1:2 gas mixture were introduced in order to promote ternary
    attachment of Ne atoms to \MP\ cations with a yield of only $\lesssim 1\,\%$.
    The depletion (single photon process) of the $\MP\cdot \textrm{Ne}$ complex was then 
    recorded as a function of wavelength.
\paragraph{IR Multi-Photon Dissociation (IRMPD) of the primary ion \MP} These experiments 
    were performed at a trap temperature of about $200\,\textrm{K}$. He gas was introduced 
    only to aid ion thermalisation and was removed prior to irradiation. The FELIX laser 
    beam was focussed into the trap region to facilitate multi-photon absorption processes. 
    The typical irradiation time to record the spectrum is $3\,\textrm{s}$ and the major 
    fragment corresponds 
    to the loss  of acetylene, \CCHH\ (cf. Table~S1 of SI). The number of produced fragment 
    ions at $m/z=189$ is therefore recorded as a function of wavelength.

The FELIX laser used for the experiment delivers up to 30~mJ into the 22 
pole trap in a single macropulse (about $5\,\mu{\textrm{s}}$ long), at a $10\,\textrm{Hz}$ 
repetition rate. The macropulse is composed of micropulses with energy in the 
$10\,\mu{\textrm{J}}$ range and with a typical length in the picosecond range. 
A spectral bandwidth of approximately $\sigma=0.5\%$ was used.

\begin{figure}[h]
\centering
  \includegraphics[]{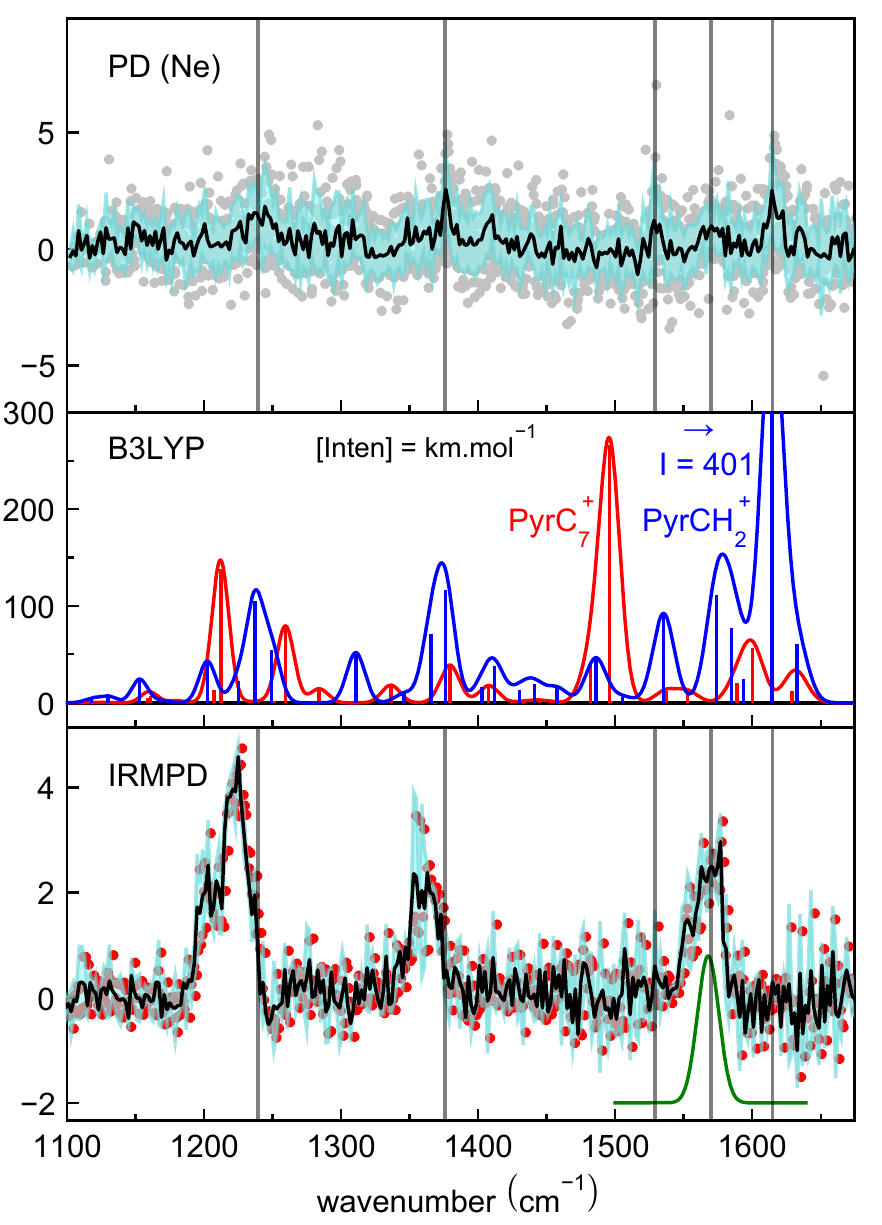}
  \caption{
  IR spectrum of \MP\ ($m/z=215$).
  Top panel -- IR pre-dissociation spectrum of the molecular ion tagged with Ne.
  Middle panel -- stick spectrum corresponding to the scaled DFT  harmonic spectra for the
  two isomers \pyrch\ (blue) and \pyrc\ (red) with intensities in km.mol$^{-1}$.
  Convoluted  spectra with $\sigma=0.5\%$ BW are provided for comparison with the experiment.
  Bottom panel -- IR multi-photon dissociation spectrum obtained by recording the 
  fragment at $m/z=189$ (loss of \CCHH). Green line represents the approximate laser line profile.
  For experimental spectra, intensities are in 
  arbitrary units and vertical lines correspond to the identified bands in the PD spectrum.
  (cf. Table~\ref{t:exp}).
    }\label{f:exp}
\end{figure}

\subsection{Computational}
Two computational approaches were used in order to i) predict the IR harmonic spectra and ii)
gain insight into the dissociation dynamics during the IRMPD process.
\paragraph{Static DFT calculations} These were carried out in order to obtain the theoretical IR 
harmonic spectra of  the two isomers of \MP. In a recent computational study of the
isomerisation reactions of \bzium\ isomers, Kharnaior et al. \cite{Kharnaior2016} have
concluded that the B3LYP exchange-correlation functional \citep{b3lyp}
provides good results for molecules of similar size and structure.
We therefore optimised the geometries of the two isomers of \MP\ using the B3LYP hybrid 
functional in conjunction with a 6-31G(d,p) basis set \citep{basisset}. 
These were found to be quasi degenerate at this level, with \pyrc\ being more stable than 
\pyrch\ by $0.5~\textrm{kJ}.\textrm{mol$^{-1}$}$ when including zero-point energy corrections.
The harmonic spectra were obtained by full 
diagonalization of the weighted Hessian matrix.
The calculations were achieved with the Gaussian09 suite of programs \citep{Gaussian2009}. 
\paragraph{Dissociation dynamics} PAHs with high enough internal energy may undergo many possible 
isomerisation reactions \cite{rev_diss_pah}, among them H migrations \cite{Georges1}, external 
ring openings and formation of 5-membered rings \cite{Georges2}. This leads to many different 
possible dissociation paths, and an exhaustive study of reactions mechanisms with DFT becomes 
very fastidious. A first insight into reaction kinetics, branching ratios and mechanisms can 
be obtained from molecular dynamics (MD) simulations in which the electronic structure is 
described on-the-fly at the Self-Consistent-Charge Density Functional based Tight Binding 
(SCC-DFTB) level of theory \cite{elstner98} (MD/SCC-DFTB), as shown in our previous work 
\cite{philtrans}.  To get insight  into isomerisation and dissociation mechanisms of the 
two isomers of \MP, we ran extensive simulations following the procedure detailed in 
Section 2 of \cite{philtrans}. We also performed calculations on the regular pyrene 
cation as a reference system.
For each molecular ion of interest, 720 simulations of $500\,\textrm{ps}$ 
($\delta\textrm{t}=0.1\,\textrm{fs}$)  were achieved with random initial velocities in the 
(NVE) ensemble, at $19\,\textrm{eV}$ (lowest energy 
necessary to observe a dissociation event) and 24 \,eV (to obtain significant branching ratios).
%  so as to reach a given internal energy, that was taken at 19 and 24 \,eV, corresponding 
%respectively  to the lowest energy necessary to observed some dissociation within 500\,ps, 
%and to a typical energy to obtain quantitative branching ratios.  
The drawback of such simulations is their time limitation (less than 1\,ns for a system of 
$\sim$30 atoms), making the use of high internal energy mandatory to observe a sufficient
number of events. The calculations were achieved with the deMonNano package \cite{demon}.\\

\section{Results and Discussion}

\subsection{Infrared spectra}
%\subsubsection{IR-PD spectra ?}
Figure~\ref{f:exp} shows the two experimental spectra that were recorded, the IR-PD spectrum
in the top panel and the IRMPD spectrum in the bottom panel. The two experimental spectra
are rather different. The IR-PD spectrum exhibits several weak bands, whereas the IRMPD
spectrum depicts 3 strong bands, which fall close to some of the bands in the IR-PD spectrum. 
Further insight into the band assignment comes from the comparison with the computed DFT harmonic
IR absorption spectra  of the two isomers, \pyrch\  and \pyrc,
which are reported in the middle panel. %In a recent computational study of the
%isomerisation reactions of \bzium\ isomers, Kharnaior et al. \cite{Kharnaior2016} have
%concluded that the B3LYP exchange-correlation functional \citep{b3lyp}
%provides good results for molecules of similar size and structure.
%We therefore optimised the geometries of the two isomers of \MP\ using the B3LYP hybrid 
%functional in conjunction with a 6-31G(d,p) basis set \citep{basisset}. The calculations 
%were achieved with the Gaussian09 suite of programs \citep{Gaussian2009}. 
Both isomers were
found to have an $^1$A' ground-state.  All band frequencies were multiplied by a 
unique scaling factor of 0.975 obtained by adjusting the position of the  most intense band
in the calculated spectrum of \pyrch\ at 1656~\CM\ to the  band located at 1615~\CM\ in 
the IR-PD spectrum. This factor is in accordance with those used previously for similar 
systems \cite{Zins09}. 
%We would like to note here that using Ne as a tagging agent generally introduces 
%only small shifts in vibrational band positions compared to the bare ion \cite{Solca2001}.
All IR bands in the $1100-1670~\CM$ region involve C-C stretching and in-plane C-H bending modes 
(a' symmetry; Figures~S2 and S3  in the SI for more information on the assignment of 
vibrational bands). They differ significantly between the two isomers.
% The spectra of the two isomers exhibit significant differences with overall more 
%intense bands for \pyrch.

Although the obtained IR-PD spectrum is rather noisy due to the difficulty of efficiently tagging
the \MP\ cation with Ne,
%(He has been tried as well, with yields even lower than those for Ne),
the comparison with the DFT spectra allows us to identify \pyrch\ in
these experiments. %{\bf: CJ: I have removed as "the dominant isomer" since the band intensities 
%of PyrC7+ are lower}. 
This assignment is based on at least 5 bands as listed in Table~\ref{t:exp}. There is no evidence
for characteristic signatures of \pyrc, but the low signal-to-noise ratio of the spectrum combined
with the expected lower band intensities of \pyrc\ compared to \pyrch\ (cf. Table~\ref{t:exp}) 
are not in favour of detecting this ion.
In the case of the IRMPD spectrum, we also observe three bands that could be attributed to \pyrch.
This conclusion is however challenged by the fact that the band measured at 1615~\CM\ in the PD 
experiments, which is the strongest in the theoretical spectrum of \pyrch, has no counterpart in
the IRMPD spectrum. The band at about 1530~\CM\ is also missing.
On the other hand, none of the predicted bands of \pyrc\ are observed in our IRMPD spectrum, 
which raises the question of the species that are traced by this latter spectrum.

% neither in the IR-PD nor in the IRMPD spectrum. In particular, the strong absorption calculated
% at 1496~\CM\ is absent, which strongly suggests that \pyrch\ is indeed the main isomer formed in 
% our experiments.

It is interesting to compare our results with previous IRMPD spectra recorded on substituted 
benzylium (Bz$^+$) and tropylium (Tr$^+$) cations. Zins et al. \cite{Zins09} compared the IRMPD 
spectrum of the tert-butyl derivative
of \bzium with B3LYP calculations using the same scaling factor of 0.975 as in our work.
They concluded that only Bz$^+$ can be observed. The characteristic band of the Tr$^+$ 
isomer at 1506~\CM\ was not observed but the authors argued that this is a sensitivity matter, 
considering that the computed absorption intensity of Tr$^+$ is a factor of $\sim$10 weaker 
than that of the strongest band of Bz$^+$. The IRMPD spectra of methyl-substituted species were
recorded by Chiavarino et al. \cite{Chiavarino2012}, showing that the major band of Bz$^+$ 
and Tr$^+$ species falls at 1600 and 1480~\CM, respectively. 
Finally, Morsa et al. \cite{Morsa2014} were able to produce preferentially the derivatives of 
Bz$^+$ and Tr$^+$ isomers of the $\textrm{C}_{11}\textrm{H}_{15}^+$ 
ion by tuning the activation regime of the precursor ion. The major 
bands of Bz$^+$ and Tr$^+$ derivatives were recorded in the IRMPD spectra at 1615 and $1425\,$\CM,
respectively. All the above studies show that Bz$^+$ has a characteristic IR band at 
$\sim\,1600\,$\CM, whereas a band is expected at $\sim\,1500\,$\CM\ for Tr$^+$ species. Although our
species are significantly different from these previously studied systems, their calculated IR 
spectra also carry these characteristic features, which are neither present in the spectrum of  the 
regular neutral pyrene nor cationic pyrene (cf. Figure~S5 in the SI). This can be 
understood if one considers
the change in charge distribution from $\textrm{C}_{16}\textrm{H}_{10}^+$ to 
\pyrch\, and \pyrc\, (see Mulliken 
charge distribution on Figure~S4 in the SI) that leads to a dipole moment of 1.29\,D for both 
isomers, in the XY plane following the orientation reported in Figure~\ref{f:iso} (more specifically
along the Y axis for \pyrc). This  is expected to lead to an enhancement of the IR intensity 
for the modes whose projection in the direction of the dipole moment is significant,  which is
the case for some $\delta _{CH} + \nu _{CC} $ modes (in plane C-H bend and C-C stretch) such as
those  reported for instance  on Figure~S2 in the SI for \pyrc\ and on Figure~S3 (e)-(f) in 
the SI for  \pyrch\ . 
%and \pyrc\ as shown in Fig. ZZZ of the SI; Note: x and y axis should appear on the figure: aude 
%: the axis are on figure 1 as mentionned just above. }. % }%the charge is mainly localised  on 
%this chemical assembly/functionality (??), which induces a significant dipole moment and there 
%IR intensity.}

\begin{table}[h]
    \caption[]{
    Experimental mid-IR band positions recorded for \MP\
    and calculated positions for \pyrch\ and \pyrc\ isomers (see Figure~\ref{f:exp}) obtained
    in the present work.
%     Anharmonicity factors $\chi$ are expressed in $10^{-2}\;\CM\cdot\textrm{K}^{-1}$,
    The intensities corresponding to the maximum of the convoluted spectra are also reported.
    %(brackets: number of  contributing transitions within $\pm 15\;\CM$).
    %All wavenumbers $\nu$ are in \CM, numbers in parentheses represent the approximate 
    %experimental linewidth (FWHM) of the strongest line component. 
    }\label{t:exp}
    \newcolumntype{Y}{>{\centering\arraybackslash}X}
    \begin{tabularx}{\columnwidth}{cYYY}
        \toprule
        \multicolumn{2}{c}{Experiment} & \multicolumn{2}{c}{Calculations} \\
        $\nu$  & $\nu$ & $\nu$ & Inten. \\
        (cm$^{-1}$)  & (cm$^{-1}$)  & (cm$^{-1}$) & ($\textrm{km}\cdot\textrm{mol}^{-1}$)\\
        IR-PD & IRMPD & \multicolumn{2}{c}{DFT-B3LYP} \\
        \midrule
        \multicolumn{4}{c}{\pyrch}\\
        1240(25) & 1223(21) & 1236  & 117\\     %ir -pd originally 1239.5
        1376(9)  & 1362(23) & 1376  & 144\\
        1529(12) & $-$      & 1535  & 92\\
        1570(21) & 1568(26) & 1574  & 154\\
        1615(16) & $-$      & ~1615* & 407\\
        \midrule
        \multicolumn{4}{c}{\pyrc}\\
        $-$ & $-$  & 1212 & 148\\
        $-$ & $-$  & 1496 & 274\\
        %1615, 1567, 1536, 1382, 1247
        \bottomrule
    \end{tabularx}
    \vskip 0.2em
    {\small{\emph{Note:}
    Consult SI for the visualisation of modes. * -- mode used to determine the scaling factor.
    }}
\end{table}

\subsection{Dissociation dynamics}

\begin{figure}[h]
\centering
  \includegraphics[]{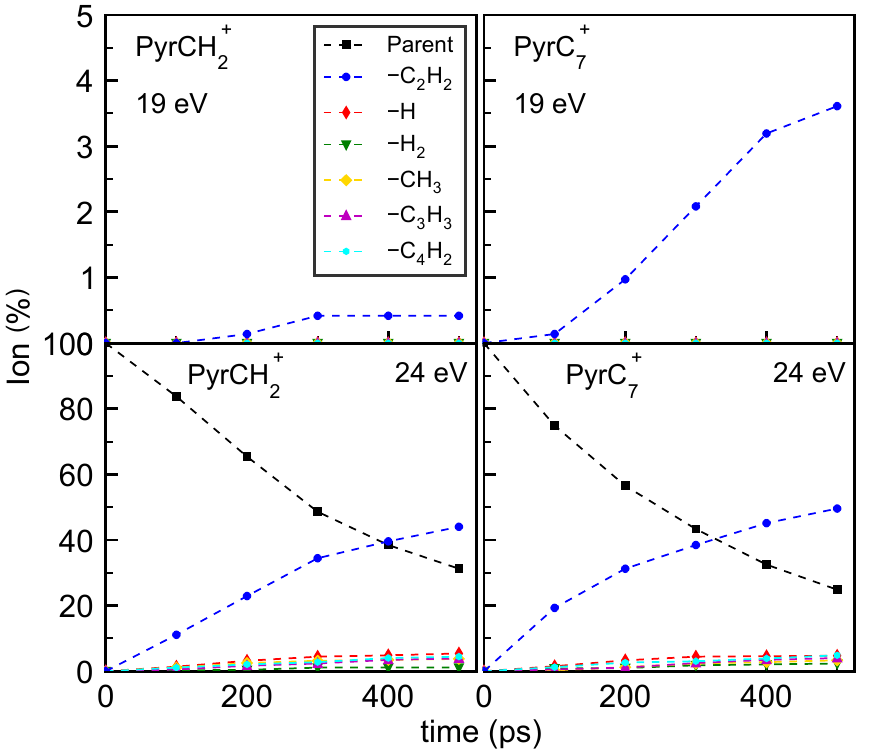}\\
   \caption{
Dissociation of \pyrch\ and \pyrc: branching  ratios as a function of time obtained 
from MD/SCC-DFTB simulations at 19 and 24\,eV.  The lines between data points are 
drawn to guide the eyes.}\label{f:kin}
\end{figure}

In order to shed more light on the measured IRMPD spectrum, we report in the following further 
insights into the dissociation dynamics of \MP\ ions. Dissociation rates, branching ratios and 
%insights into the 
structures of the products and reaction intermediates were obtained from 
MD/SCC-DFTB simulations following the procedure detailed in Section 2.2 and starting from the 
two isomers \pyrch\ and \pyrc. The branching ratios and dissociation rates are determined at 
24 and 19\,eV (cf. Figure~\ref{f:kin}). The major fragment for both isomers corresponds to 
the loss of \CCHH, H loss being second, but much lower. At both calculated energies, the 
dissociation of \pyrc\ is found to be slightly more efficient than that of \pyrch.
We compared these results with calculations on the regular pyrene cation (cf. Figure~S6 in 
the SI). For this ion, we know from comparison with photo-absorption experiments 
\cite{West2014} that our calculations overestimate the contribution of the \CCHH\ channel and 
this challenging issue is currently under investigation. Still the \CCHH/H fragmentation 
branching ratio is found to be larger for both isomers of \MP\ than for cationic pyrene. 
This indicates that \CCHH\ is a major fragmentation channel for \MP\ ions, which is also 
shown by our experimental results (cf. Table~S1 in the SI) and a previous study \citep{Kokkin2014}.

\begin{figure}[h]
\centering
  \includegraphics[width=.50\textwidth]{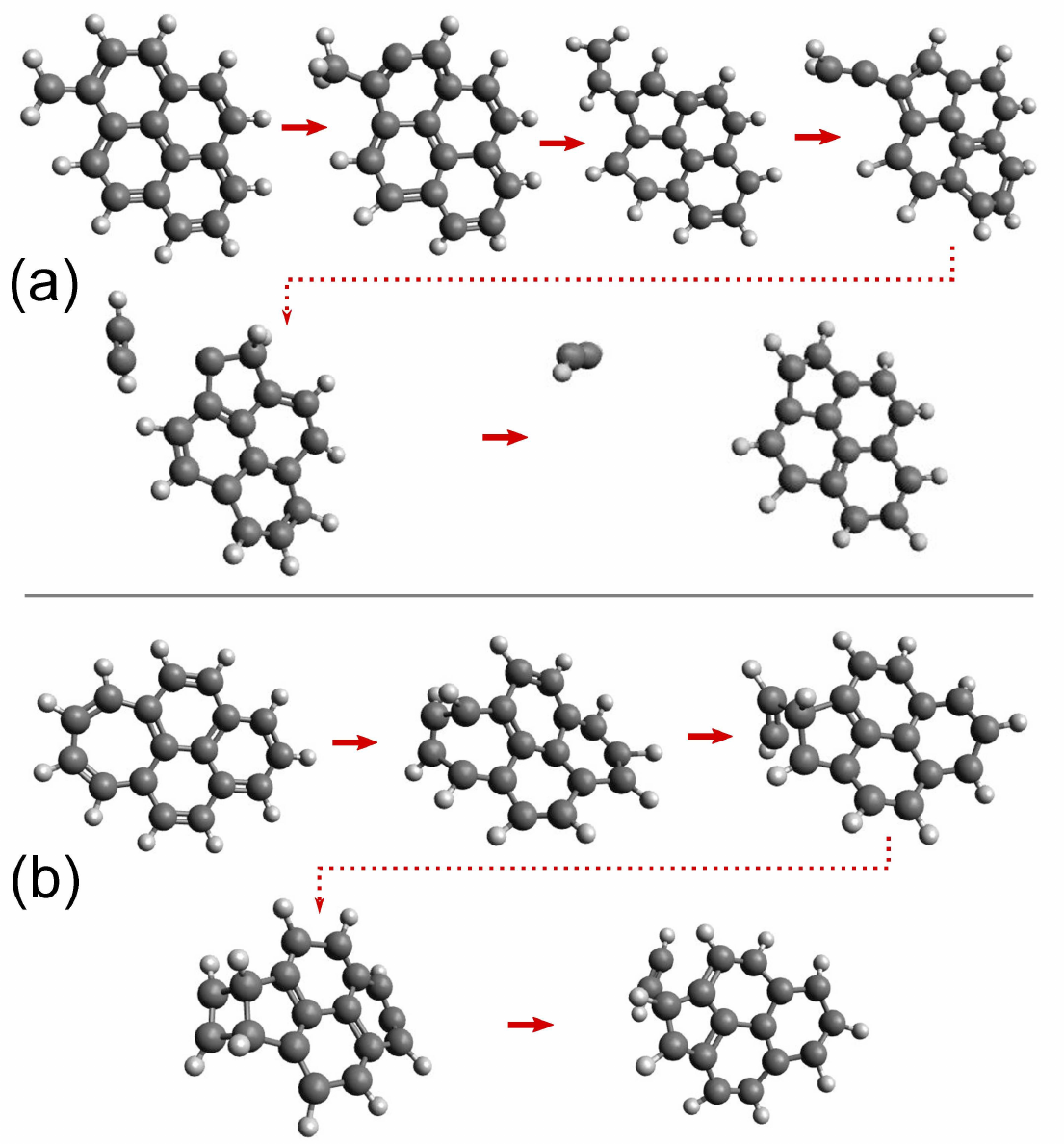}
  \caption{
 Examples of dissociation paths : snapshots extracted from 500\,ps MD/SCC-DFTB trajectories 
 run at 19\,eV  for \pyrch (a) and \pyrc (b).
    }\label{f:paths}
\end{figure}

From our simulations at 19\,eV, we could get insights into the product structures and 
possible dissociation mechanisms. Some snapshots of simulations that we estimated to be 
representative are reported in Figure~\ref{f:paths}  and in Figure~S7 in the SI for 
both \pyrch\ and  \pyrc. We found that a common structure for the dissociation product 
is the presence of a five-membered ring as earlier suggested \cite{Kokkin2014}.
The other noticeable product has undergone ring opening to form an acetylene terminal group.
This terminal structure was also observed as the final product in several simulations 
for both isomers. It is presented in the case of \pyrch\ in Figure~S7 (a) in the SI.

All paths were found to involve H migrations, leading to the formation of quite long time 
scale intermediates. H migration also leads to the formation of sp$^3$ carbon atoms and 
weakening of the CC bond. In the case of \pyrch, one can notice the formation of a 
terminal -CH-CH$_2$ vinylidene function, which often appears as a long time intermediate 
in the simulations (cf. Figure~\ref{f:paths} (a)). On the other example reported in 
Figure~S7 (a) in the SI, we see that isomerisation into the tropylium-like isomer occurs 
following a mechanism invoked in Rapacioli et al. \cite{Rapacioli2015} (out of plane 
deformation of the PAH, CH$_2$ insertion and H migration). In the case of \pyrc, one can 
observe the direct formation of a 5-membered ring (through a [5+4] intermediate, see 
first structure at the bottom of Figure~\ref{f:paths} (b)) from the 7-membered cycle, 
followed by \CCHH\ loss. In the other selected path in Figure~S7 (b) in the SI, one can 
see the variety of possible isomerisation reactions involving H migration, external ring 
opening and formation of acetylene terminal groups and five-membered rings. Interestingly, 
only the tropylium cycle in \pyrc and the  C$_6$-CH$_2$  part of \pyrch are altered, 
showing their enhanced fragility with respect to the rest of the carbon skeleton.
% \textcolor{blue}{ aromatic?aude : je ne sais pas s'il faut mettre ça car le cycle detruit 
%contribue largement a l'aromaticite au départ d'après mes connaissances (autant que le 
%reste du squelette je dirais), mais au moins deux des cycles restants devraient être aromatiques }.

Overall, the results of these simulations can help us to rationalise the IRMPD spectrum. 
The simulations suggest a lower but comparable internal energy to dissociate \pyrch\ and 
\pyrc\ as well as the pyrene cation. The dissociation threshold for the pyrene cation has
been observed to be around 9\,eV in other ion trap experiments using synchrotron 
radiation \cite{Zhen2016}, corresponding to a micro-canonical temperature of 2200\,K. We 
can therefore use this value as a first estimate of the internal energy required to 
dissociate \MP. With this amount of internal energy, the system will be able to overcome 
the \pyrch\ to \pyrc\ isomerisation barrier of $3.5-4.0\,\textrm{eV}$ \cite{Rapacioli2015}. Our 
simulations also suggest that the heated molecular ion can experience many isomeric forms 
on its way to dissociation involving not only the two lowest-energy isomers but also 
higher-energy ones. 
For IRMPD to occur, the irradiated ion should absorb continuously the IR laser irradiation 
to reach high enough internal energy/ temperature to dissociate. Isomerisation will affect 
the IR absorption spectrum. Figure~S8 of the SI aims at illustrating this effect by mixing 
the harmonic spectrum of \pyrch\ with those of the 3 isomers involved in the calculated 
dissociation path presented in Figure~\ref{f:paths}, panel (a). Although this spectrum 
cannot be compared directly to the IRMPD spectrum, it shows the spectral ranges with higher 
global absorption cross-section for the mixture of isomers. These spectral ranges are likely 
to have higher heating efficiency upon laser irradiation and therefore a higher IRMPD signal. 
Although this constructed global spectrum shows 3 comparably intense bands, 
it still does not explain the missing 1615 \CM\ band in the experimental IRMPD spectrum.
Anharmonicity is another effect that can affect the IR spectrum.
Rapacioli et al. \cite{Rapacioli2015} calculated the IR spectra of  
\pyrch\ and \pyrc\ as a function of temperature up to 2000\,K. The spectra exhibit significant 
band shifts and broadening with temperature. Hence, the recorded IRMPD spectrum is expected 
to reflect the net efficiency of heating the ion while the absorption IR spectrum is 
modified by both isomerisation and anharmonic effects, making its interpretation very tricky.
The impact of isomerisation during the IRMPD process has been previously discussed in the 
case of C$_7$H$_9^+$ \cite{Schroder2006}. We note that this issue has not been reported in the 
IRMPD studies on substituted benzylium (Bz$^+$) and tropylium (Tr$^+$) cations that are 
described in the previous section. The reason is that, for these systems, the involved 
dissociation energies were much lower making complex isomerisation pathways less competitive.

% A possible explanation is that this band is strongly anharmonic,
% which indeed has been predicted by Rapacioli et al. in their finite temperature molecular
% dynamics simulations (cf. \cite{Rapacioli2015} and the respective anharmonicity values, 
% $\chi$, listed in Tab.~\ref{t:exp}). One has to recall that the IRMPD signal reflects 
% the efficiency of absorbing multiple photons within the laser bandwidth, which depends 
% on the evolution of the band position and width, while the internal temperature is 
% increased.
% A similar effect can be invoked for the other missing band at about 1530~\CM.

\section{Conclusion}

In summary, this work demonstrates the benefit of the combination of a widely tunable
IR laser source (FELIX) with cold ions prepared in a 22-pole ion trap in probing the 
role of isomerisation in the photophysics of PAHs. 
In the specific case of \MP, we showed that the IRMPD spectrum is difficult to analyse 
and can only bring limited information.
This spectrum lacks the characteristic band around $1600\;\CM$, which 
has been observed in previous IRMPD spectra of benzylium-type species. Possible reasons 
include anharmonic effects and isomerization while heating the trapped ions, as suggested 
by molecular dynamics simulations. Further experimental work either on the anharmonic 
spectra or on the dissociation dynamics of \pyrch\ would help to solve this issue.
On the opposite, a clear relation between the one photon absorption 
bands of the IR-PD experiment and DFT predictions unravels the presence of the  
\pyrch\ isomer in our experimental conditions. The \pyrc\ isomer was not observed 
due to the limited signal-to-noise ratio. Thus, the presence of this isomer at 
formation cannot be excluded.

% We found that \pyrch\ is the main isomer in our experimental conditions.
% Still, one has to remind that these ions were prepared by electron impact ionization
% in our study. In order to verify that \pyrch\ would also be the main product in
% photodissociation experiments, it would be desirable to combine structural 
% determination with the excitation in the  UV-VIS range, i.e. to first produce the 
% photoproduct of the 1-methylpyrene cation and subsequently determine its structure 
% by the IR action spectroscopy described in this work. 

\section*{Acknowledgments}

We greatly appreciate the experimental support provided by the FELIX team.
The research leading to this result is supported by the European Research 
Council under the European Union's Seventh Framework Programme ERC-2013-SyG, 
Grant Agreement n. 610256 NANOCOSMOS. We acknowledge support from the project
CALIPSOplus under the Grant Agreement 730872 from the EU Framework
Programme for Research and Innovation HORIZON 2020. S. B. and St. S. also 
acknowledge support from DFG SPP 1573 grant BR 4287/1-2.
P. J. and the operation of the 22-pole ion trap were partially funded by the 
DFG via the Ger\"atezentrum ``Cologne Center for Terahertz Spectroscopy'' 
(DFG SCHL 341/15-1). G. W. is supported by the H2020-MSCA-ITN-2016 Program 
(EUROPAH project, G. A. 722346).
A. S. would like to thank the computing facility CALMIP for generous 
allocation of computer resources. Finally, we thank the anonymous referees 
for helping improving this manuscript, in particular the analysis of the 
IRMPD spectrum.

\section*{Supplementary material}

Supplementray material can be found online at: \url{https://doi.org/10.1016/j.cplett.2018.03.028}.

\section*{References}

\bibliography{m-pyrene}

\end{document}